\documentclass[twocolumn,groupedaddress]{revtex4-1}
\usepackage{graphicx}
\usepackage{dcolumn}
\usepackage{bm}
\usepackage{subfigure}
\usepackage{color}
\usepackage{amsmath}

\newcommand {\BU}{Mechanical Engineering Department and the Photonics Center, Boston University, Boston, Massachusetts 02215, USA}

\begin{document}

\title{Noninvasive Measurement of the Pressure Distribution \\ in a Deformable Micro-Channel}

\author{O. Ozsun}
\affiliation{\BU}

\author{V. Yakhot}
\affiliation{\BU}

\author{K. L. Ekinci}
\email[e-mail:] {ekinci@bu.edu}
\affiliation{\BU}

\begin{abstract}
\emph{Direct and noninvasive} measurement of the pressure distribution in test sections of a micro-channel is a challenging, if not an impossible, task. Here, we  present an  analytical method for extracting the pressure distribution in a deformable micro-channel under flow. Our method is based on a measurement of the channel deflection profile as a function of applied \emph{hydrostatic} pressure; this initial measurement generates ``constitutive curves" for the deformable channel. The deflection profile under flow is then matched to the constitutive curves, providing the \emph{hydrodynamic} pressure distribution. The method is validated by measurements on planar micro-fluidic channels against analytic and numerical models. The accuracy here is independent of the nature of the wall deformations and is not degraded  even in the limit of large  deflections,  $\zeta_{\rm{max}}/2h_{0}= {\cal{O}}(1)$, with $\zeta_{\rm{max}}$ and $2h_0$ being  the maximum deflection and the unperturbed  height  of the channel, respectively. We discuss  possible applications of the method in characterizing  micro-flows, including those in biological systems.
\end{abstract}

\maketitle

\section{Introduction}
Since the early experiments of Poiseuille \cite{skalak} more than two centuries ago, the craft of  measuring flow fields in tubes and pipes has been perfected.  Even so, the resolution limits of these exquisite experimental probes --- such as the pitot tube \cite{McKeon} or the hot wire anemometer \cite{anemometry} --- are quickly being  approached, given recent advances in micron and nanometer-scale technologies. One frequently encounters micro- \cite{stone_review, Popel} and nano-flows \cite{nanofluidics}, which come with smaller length scales \cite{Lissandrello} and shorter time scales \cite{ekinciLoC10,ekincireview} than can be resolved by the commonly available  probes. For instance, in a pressure-driven micro-flow, one must insert micron-scale pressure transducers in test sections in order to determine the local pressure drops \cite{Hardy, Akbarian, Kohl, Orth, Song, Srivastava}. As the size of a probe becomes comparable to or even bigger than the flow scale itself, measurement of the distribution of flow fields becomes problematic.

Although the tools of macroscopic fluid mechanics may not easily be scaled down, the materials and techniques of micro-fluidics  offer unique measurement approaches. Most micro-channels in lab-on-chip systems, for instance, are made up of  flexible materials \cite{Whitesides, Gervais}. This provides the possibility of probing a flow by monitoring the response of the confining  micro-channel to the flow. In other words, the local (position-dependent) deflection $\zeta$ of the \emph{deformable} walls of a micro-channel may enable the accurate determination of the pressure field (or the velocity field) under flow. The challenge in this approach, of course, is  characterizing the interactions between a deformable body and a flow \cite{Shelley, Hosoi, Holmes}. This is not a simple task, especially  in the limit of large deflections. To accurately predict a flow bounded by a deformable wall, one  needs to determine  the hydrodynamic fields as well as the wall deformations \emph{consistently}. This requires solving coupled fluid-structure equations \cite{Pedley-Carpenter, Pedley},  often in situations where constitutive relations or parameters  describing fluid-structure  interactions  are not available.  Even if these relations and parameters are known,  numerical approaches are often expensive.

To make the above-discussion more concrete, let us consider a  steady pressure-driven flow between an infinite rigid plate at $y=0$ and a deformable wall at $y=2h_0+\zeta(x)$, where $\zeta(x)$ is the local deflection of the deformable wall due to  the local pressure $p(x)$ as shown in figure~1(a).  The equations for incompressible steady flow  ($\partial_{x}u+\partial_{y}v=0$),
\begin{eqnarray}\label{NS-channel}
u\partial_{x}u+v\partial_{y}u=-\partial_{x}p+\nu({\partial_{x}}^{2}+{\partial_{y}}^{2})u,\nonumber\\
u\partial_{x}v+v\partial_{y}v=-\partial_{y}p+\nu({\partial_{x}}^{2}+{\partial_{y}}^{2})v,
\end{eqnarray}
are to be solved subject to  boundary conditions $u|_{B}=v|_{B}=0$. All the variables in (\ref{NS-channel}) have their usual meanings [see figure 1(a)], with $\nu$ being the kinematic viscosity.  In general, solution to these equations,  accounting for inlet and outlet effects are impossibly difficult. However, if the channel is long such that  ${{2h_{0}+\zeta_{\rm{max}}}\over l}\ll1$, where $\zeta_{\rm{max}}$ is the maximum deflection and $l$ is the length of the channel, we can write the local solution for the average velocity $\bar u(x)$ as
\begin{equation}\label{analytic_def_chnl}
\bar u(x) \approx {1 \over {12\eta }}{\left[ {2{h_0} + \zeta \left( {p(x)} \right)} \right]^2}{\partial _x}p.
\end{equation}
Here, $\eta$ is the dynamic viscosity, and $\zeta(x)=\zeta \left( {p(x)} \right)$ is a \emph{local} constitutive relation, which determines the dependence of the wall deflection $\zeta(x)$ on $p(x)$. No particular form for this dependence (e.g., elastic) is assumed \emph{a priori}.   In order to find the flow rate and the wall stresses, we need accurate information on  $p(x)$,  $\zeta(x)$, and the constitutive relation $\zeta(x)=\zeta \left( {p(x)} \right)$. If the flow rate is given, the problem becomes somewhat simplified, but still remains rather complex to be handled numerically or analytically.

In this manuscript, we describe a method to close (\ref{NS-channel}) in a deformable channel using \emph{independent static} measurements of $\zeta=\zeta \left( {p} \right)$. Using this method, we  extract the pressure distribution in a planar channel flow and validate our measurements  against the analytic approximation in (\ref{analytic_def_chnl}) for a long channel. Our method does not depend upon the particulars of the local constitutive relation $\zeta(x)=\zeta \left( {p(x)} \right)$. In other words, it remains independent of the nature of the wall response, providing accurate results for buckled walls and elastically  stretching walls alike.

\begin{figure*}
\centering
\includegraphics[width=5.1in]{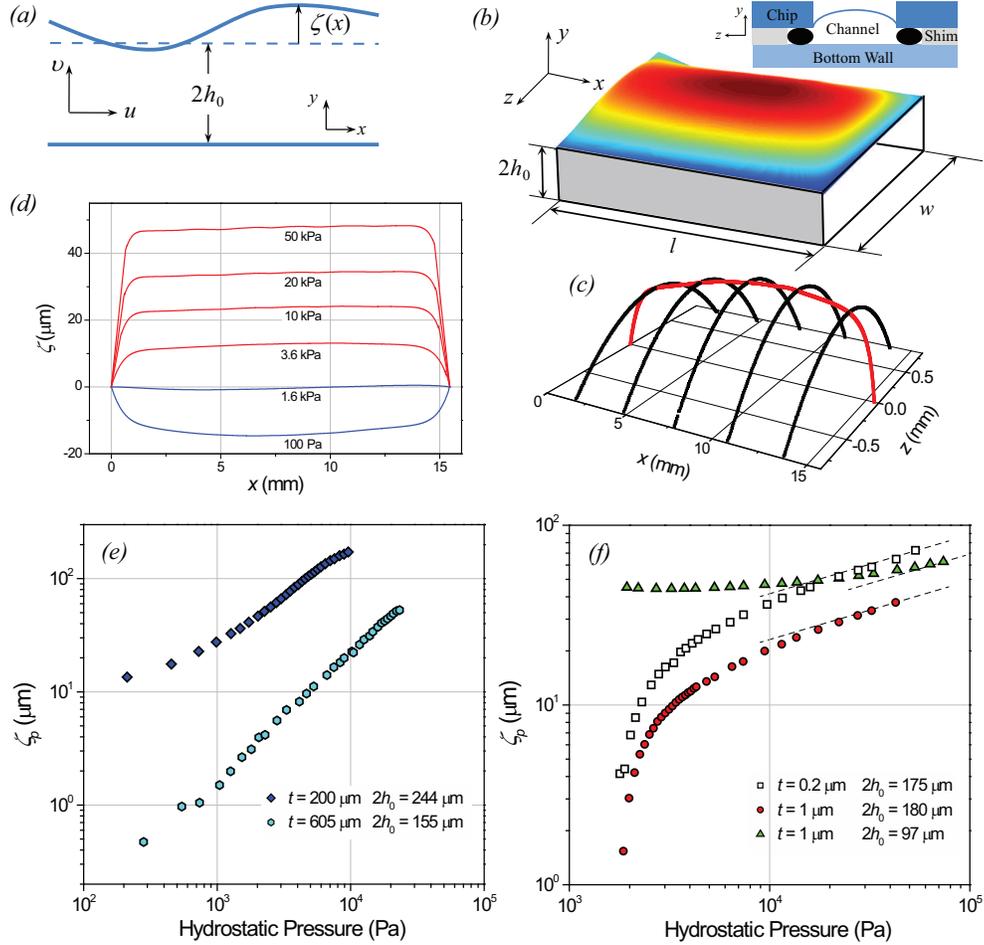}
\caption{(a) A  one-dimensional channel  with a deformable top wall.  (b) The two-dimensional deflection $\zeta_{2d}(x,z)$ of S1 ($t$=200 nm, $2h_0=175~\mu$m) under hydrostatic pressure of $p=2.3$ kPa as measured optically. The inset shows a cross-sectional view of the channel. Clamps hold  the chip (with the thin membrane at the center) and the  bottom wall together. The two walls are separated by a precision shim; an O-ring (black ovals) seals the channel.  (c) Cross-sectional line scans from the image in (b) showing the parabolic $z$-profile of the deformable wall under pressure. (d) The  one-dimensional (average) wall deflection $\zeta(x)$ at several different pressures for the same sample. The indicated values are gauge pressure values. These are the \emph{local} constitutive curves. (e) The peak deflection $\zeta_p$ of the PDMS deformable walls as a function of pressure $p$. (f)  $\zeta_p$ of the SiN deformable walls as a function of  $p$; the dashed lines are the $p^{1/3}$ asymptotes. Note that the  $\zeta_p$ and the $p$ axes do not cover the same ranges in (e) and (f). Error bars are smaller than the symbols.}
\label{Fig2}
\end{figure*}

\section{Experimental System}
To test these ideas, we have fabricated planar micro-channels with deformable walls and measured the deformations of these micro-channels using optical techniques under different conditions --- following the work of Gervais \emph{et al.}  \cite{Gervais}. Figure 1(b) is a rendering of one of our micro-channels under pressure. The inset shows how the channel is formed: a rigid bottom wall  and a deformable top wall are held together by clamps, and the channel is sealed by a gasket. The in-plane linear dimensions of the channels are $l \times w = 15.5 \times 1.7$ mm$^2$. The distance $2h_0$ between the undeflected top wall and the rigid bottom wall is set by a precision metal shim (in the range $100~\mu{\rm{m}} \lesssim 2h_0 \le 250~\mu$m); optical interferometry is employed to independently confirm the $2h_0$ values. Different materials with varying thicknesses $t$ are used to make the deformable walls. In three of the channels studied here, the deformable walls are ultrathin  silicon nitride (SiN) membranes fabricated on a thick silicon handle chip ($t\approx500~\mu$m). In the other  channels, the compliant walls are made up of thicker elastomer (PDMS) layers. Various parameters of all the micro-channels used in this study are given in table~\ref{tab:experiments}.

After the micro-channels are formed, they are connected to a standard microfluidic circuit equipped with pressure gauges. In the hydrostatic measurements, the outlet of the micro-channel is clogged, and  a water column is used to apply the desired pressure. In flow measurements, a syringe pump is inserted into the circuit to provide the flow.

\begin{table*}
  \begin{center}
\def~{\hphantom{0}}
      \begin{tabular}{lcccccc}
        Sample & Material & Wall thickness & Unperturbed height & ${\rm{Re}}$ &  Max. defl.  & Max. flow rate \\
         &  & $t$ ($\mu$m) &  $2h_{0}$ ($\mu$m) & & $\zeta_{\rm{max}}$ ($\mu$m) & $Q_{\rm{max}}$ (ml/min)  \\
        \hline
        S1 & SiN & 0.2 & 175 & 70-1200 & 33 & 70   \\
        S2 & SiN & 1 & 180 & 100-1200 & 20 & 70   \\
        S3 & SiN & 1 & 97 & 250-1300 & 37 & 70 \\
        S4 & PDMS & 200 & 244 & 200-800 & 86 & 50 \\
        S5 & PDMS & 605 & 155 & 200-900 & 25  & 50 \\
      \end{tabular}
        \caption{Parameters of the channels (first four columns), the range of Reynolds numbers (${\rm{Re}}$), and the maximum channel deflection $\zeta_{\rm{max}}$ attained under the maximum flow rate $Q_{\rm{max}}$. The ${\rm{Re}}$ is found by averaging ${\rm{Re}}_x={2Q \over {\nu [w+2h(x)]}}$  over the channel.}
        \label{tab:experiments}
    \end{center}
\end{table*}

\section{Results and Discussion}
\subsection{Hydrostatic Loading}
First, the channels are characterized under hydrostatic loading. In these experiments the inlet port is connected to a water column with the outlet clogged, and hydrostatic pressure $p$ is applied on the channel by raising the water column. The resulting position-dependent deflection field $\zeta_{2d}(x,z)$ of the compliant wall is measured  using white light interferometry \cite{SWLI} at each pressure. In figure 1(b), $\zeta_{2d}(x,z)$ of the deformable top wall of sample S1 ($t=200$ nm and $2h_0=175~\mu$m; table~\ref{tab:experiments} first row) at $p=2.3$ kPa is shown. Cross-sections along  the $x$ and $z$ axes taken from this profile are shown in figure 1(c).  Because the cross-sections are parabolic in the  $z$-direction, we define an \emph{average} or \emph{one-dimensional} wall deflection $\zeta(x)$ as
\begin{equation}\label{twothirds}
     \zeta (x) = {1 \over w}\int\limits_{ - w/2}^{ + w/2} {\zeta_{2d} (x,z)dz}  \approx {2 \over 3}\zeta_{2d} (x,z = 0).
\end{equation}
Here, $\zeta_{2d} (x,z = 0)$ is the maximum value of the parabolic cross-section, and the factor ${2 \over 3}$ comes from the integration. Similarly defined  $ \zeta(x)$ will allow us to perform a one-dimensional analysis in the hydrodynamic case. In figure 1(d), we plot $\zeta(x)$ for the same channel at several different \emph{hydrostatic} pressures, 100 Pa $\le p \le$ 50 kPa. These are the \emph{position-dependent (local) constitutive curves}. Because of the clamping stresses, the deformable wall is initially in a buckled state. At low $p$, the wall deformation remains in the  negative $y$ direction. As $p$ is increased, the wall response becomes elastic, and the wall stretches like a membrane.  Also, a small asymmetry is noticeable in  $\zeta(x)$, caused by the deformation of the silicon chip during clamping. Figure 1(e) and (f) shows the peak deflection $\zeta_{p}$, which typically occurs at $(x,z)\approx({l\over 2},0)$, as a function of $p$ for the elastomer (PDMS) and SiN walls, respectively.  Each deformable wall in figure 1(e) and (f) has a constitutive $\zeta_p$ vs. $p$ curve, determining the behavior of the \emph{entire} wall. The thin nitride walls shown in figure 1(f) obey the well-known elastic shell model at high $p$, $\zeta_p \sim p^{1/3}$ \cite{bulge}. The elastomer walls in figure 1(e) follow a different power law from the SiN ones, presumably because they are much thicker and bending dominates their deformation. There is no noticeable universality in the $\zeta_p$ vs. $p$ data --- i.e., the nature of the wall response and thus the constitutive relations are material and geometry (thickness) dependent. Our flow results below, however, remain independent of the wall response.

\subsection{Flow Measurements}
Next, we perform flow measurements in each micro-channel. The  results from all five channels are shown in figure 2(a). In the experiments, we establish a constant volumetric flow rate $Q$ through each channel using a syringe pump and  measure the pressure drop between the inlet and outlet using a macroscopic transducer.  We prefer to plot $Q$ as the independent variable because the experiments are performed by varying $Q$ and measuring the pressure drop. In all measurements, a small pressure drop occurs in the rigid inlet and outlet regions of the channel. This is because of the finite size of the connections to the macroscopic pressure transducers. Knowing the geometry of the rigid regions,  we  determine the pressure drop in these regions from flow simulations (see Appendix A for details). Subsequently, we subtract this ``parasitic pressure drop" from the measured pressure drop. In summary, $\Delta p_t$ in the plots in figure~2(a) corresponds to the \emph{corrected} pressure drop in the compliant section of the channel as measured by a macroscopic transducer (hence, the subscript ``t"). Figure 2(b) shows the channel deflection $\zeta(x)$ at several different flow rates for S1  ($t=200$ nm and $2h_0= 175~\mu$m). Returning to table~\ref{tab:experiments}, we now clarify that $\zeta_{\rm{max}}$ corresponds to the maximum deflection of the channel at the highest applied flow rate.

\begin{figure}
\centering
\includegraphics[width=3.2in]{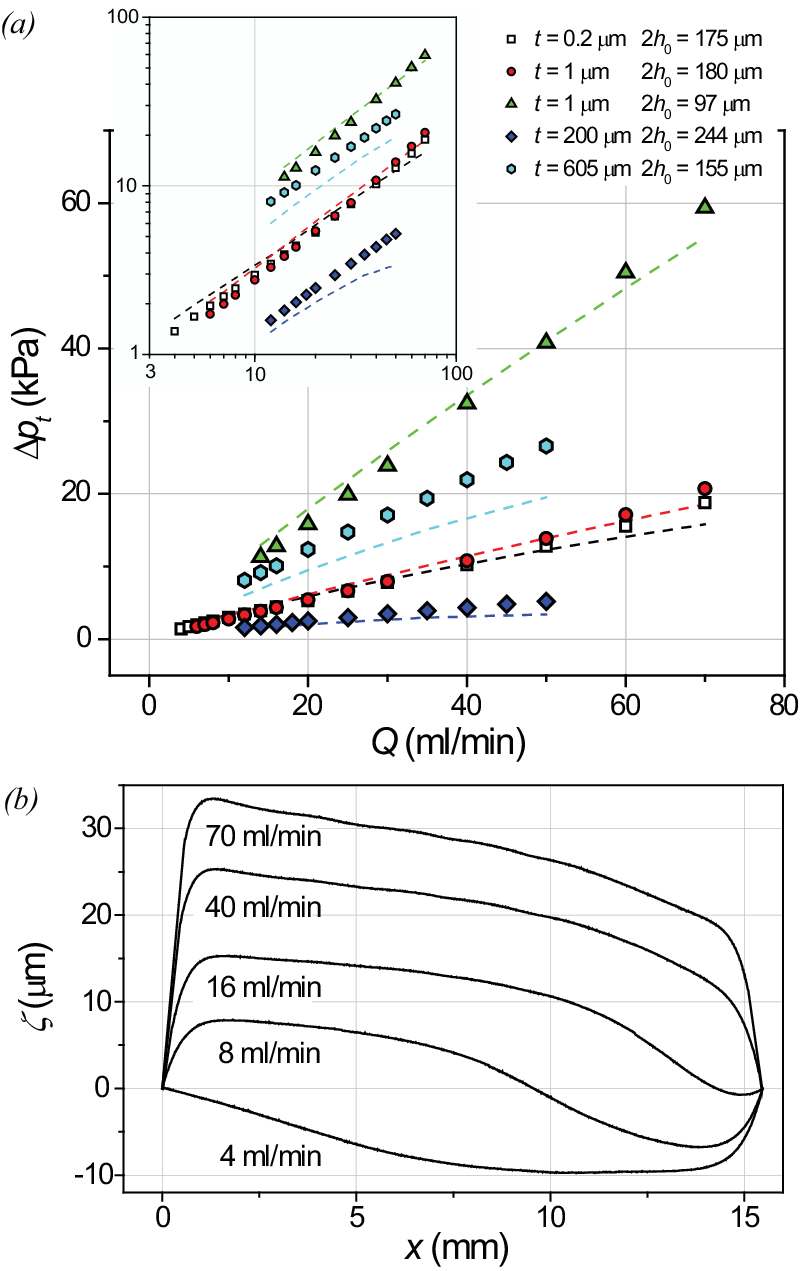}
\caption{ (a) The pressure drop $\Delta p_t$  in the compliant sections of the micro-channels as a function of flow rate $Q$. Error bars are smaller than the symbol sizes. The dotted lines show fits based on the hydraulic resistance of the micro-channel. The inset shows a double logarithmic plot of the same data. (b) The  deflection  profile $\zeta (x)$ of S1 ($t$=200 nm, $2h_0=175~\mu$m)  at different $Q$. The  profile is no longer uniform [cf. figure~1(d)] because of the position-dependent pressure $p(x)$ in the channel.}
\label{Fig1}
\end{figure}
\subsection{Simple Fits}
Before we present our method for analyzing the flow, we attempt to fit the experimental $\Delta p_t$ vs. $Q$ data to the theory described above in Eqs.~(\ref{NS-channel}-\ref{analytic_def_chnl}). Because our channels have a finite width $w$, the result in (\ref{analytic_def_chnl}) must be modified slightly. The simplest approach is to  use a linear approximation for the local pressure drop based on the hydraulic resistance per unit length, $r(x)$, of the channel.  In a long channel at low Reynolds number,   ${\partial _x}p\approx Q r(x)$. The total pressure drop  between the inlet and outlet can then be found as $\approx Q \int\limits_0^l {r(x)dx}$ \cite{batchelor}.  In our analysis, we approximate our channel as a  channel of rectangular cross-section of $w \times 2h(x)$, where $2h(x)=2h_0+\zeta(x)$. Then \cite{bruss},
\begin{equation}\label{hyd_res_unit_length}
    r(x) \approx \left( {{1 \over {1 - 0.63\left( {{{2h(x)} \over w}} \right)}}} \right){{12\eta } \over {w{{\left[ {2h(x)} \right]}^3}}}.
\end{equation}
With the $2h(x)$ data available from optical measurements, we calculate $r(x)$ and integrate it along the length of the channel for all flow rates to find the pressure drop. The calculated $\Delta p_t$ are shown in figure 2 as dashed lines. It is difficult to determine the source of the disagreement between the data and the fits in some cases. The flow in the inlet and outlet regions may still be contributing to the error --- even after subtraction. Another source of error may be the boundary between the compliant and rigid regions of the channel.


\begin{figure*}
\centering
\includegraphics[width=5.1in]{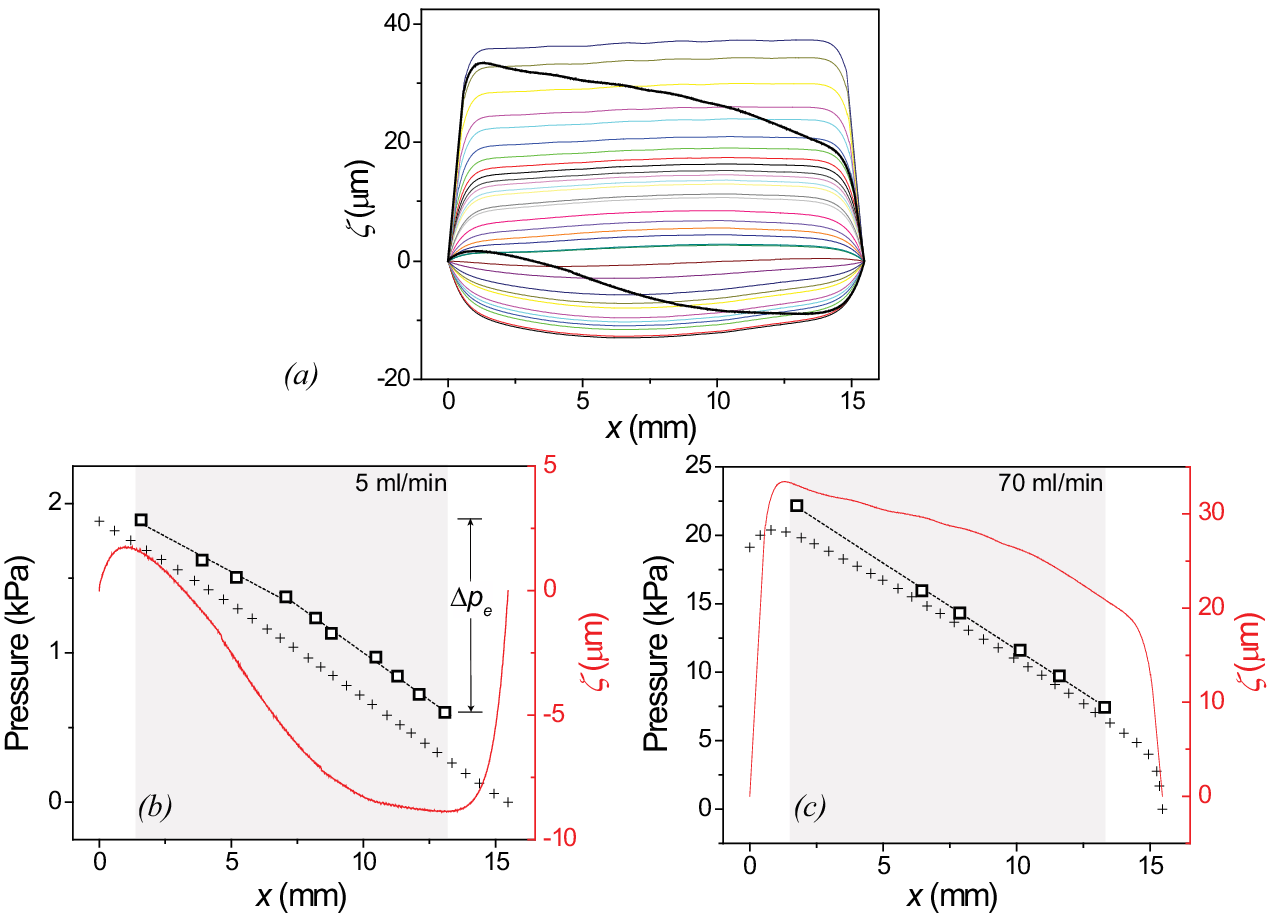}
\caption{Determining the pressure gradient inside a microchannel from hydrostatic measurements (S1, $t$=200 nm, $2h_0=175~\mu$m). (a) The channel deflection profiles under two flow rates (upper thick line, $Q=70$ ml/min; lower thick line, $Q=5$ ml/min) are overlaid on top of the  deflection profiles taken under different hydrostatic pressures [cf. figure 1(d) and figure 2(b)]. (b) The data points correspond to the pressure distribution $p(x)$  in the test section (shaded region) for a flow rate of $Q=5$ ml/min. To obtain $p(x)$, the pressure values at the intersection points in (a) are plotted as a function of $x$. A \emph{nonlinearity} is noticeable in $p(x)$ at $x\approx 7.5$ mm, where the slope of the linear fit changes. (c) Similarly determined $p(x)$ for $Q=70$ ml/min. The dashed line is a simple linear fit. The solid (red) lines in (b) and (c) are the deflection profiles of the channel at the given flow rates. The error bars in the data are estimated to be smaller than the symbol size. Also plotted (+) are results from simple flow simulations.}
\label{Fig3}
\end{figure*}
\subsection{Measurement of the Pressure Distribution}
We now turn to the main point of this manuscript. Our method is illustrated in figure 3. The curves in the background in figure 3(a) are the now-familiar $\zeta (x)$ curves of S1 ($t=200$ nm and $2h_0=175~\mu$m) under different \emph{hydrostatic}  pressures [cf. figure 1(d)]. These serve as the constitutive curves. On top of these \emph{hydrostatic} profiles, we overlay two different \emph{hydrodynamic} profiles (thicker lines) at flow rates of $Q=5$ ml/min and $Q=70$ ml/min. The assumption here is that, under equilibrium, $\zeta (x)$ only depends upon $p(x)$, providing us with the constitutive relation $\zeta(x)=\zeta \left( {p(x)} \right)$. We determine the positions where the dynamic profiles intersect with the static profiles and read out the pressure values for each intersection position. In figure~3(b) and (c), we plot these read-out pressure values using symbols  as a function of position for the two different flow rates ; solid (red) lines are the  deflection profiles; also plotted (+) are results from simple flow simulations (see below for further discussion). In figure 3(b), a  noticeable deviation from a linear pressure distribution is present, as captured by the two dashed line segments with different slopes. The $p(x)$ in figure 3(c) can be approximated well by a linear fit (dashed line) to within our  resolution. In the region near the boundaries ($x=0$ mm and $x=15.7$ mm), where significant pressure gradients must be present, it is not possible to obtain pressure readings. Thus, the test section is the shaded regions in figure 3(b) and (c) away from the boundaries.  We confirm that similar behavior is observed in all measurements on different channels.

Finally, we show that what is found above is indeed the pressure distribution in the channel. First, we turn to the simple flow simulation results (shown by +) in figure 3(b) and (c). Here, we take a two-dimensional channel with two \emph{rigid} walls, with the top one having the experimentally-measured profile $\zeta(x)$ and the bottom one being flat. We prescribe the velocity $u$ at the inlet based on the experimental $Q$ value. We then calculate the pressure distribution in the channel with the outlet pressure set to zero (see Appendix A for more details). In figure 3(b), a small nonlinearity qualitatively similar to that observed in the experiment is noticeable. Between the experiment and the simulation, there is a small but constant pressure difference ($\sim300$ Pa), which probably mostly comes form the non-zero outlet pressure in the experiment. In figure 3(c), we notice a constant pressure difference ($\sim500$ Pa) between experiment and simulation as well; in addition, there is a larger pressure difference towards the inlet. The excess pressure observed in the experiment is probably the pressure that is needed to keep the deformable wall stretched --- as the deformability of the wall is completely ignored in the simulation. The wall is stretched more towards the inlet --- hence the larger pressure difference. (We estimate that this tension is \emph{not} present in the \emph{buckled} wall of figure 3(b).) Overall, the agreement is quite satisfactory.

We can further validate the extracted pressure drop $\Delta p_e $ across the (entire) deformable test section against the analytical approximation in (\ref{analytic_def_chnl}). Our method provides $\Delta p_e $ directly for each flow rate. We illustrate this in figure 3(b): we  take the high and low pressure values at the beginning and end of the test section, and calculate the difference to find   $\Delta p_e$, i.e., $\Delta p_e = p({x \approx 1.6~{\rm{ mm}}})-p({x \approx 13~{\rm{ mm}}})$.  Against this $\Delta p_e$ value, we plot $QR$, where $R$ is the  hydraulic resistance of the channel for \emph{only}  the region where the pressure drop is determined, i.e., the hydraulic resistance of the test section. For the data in figure 3(b), for instance, $ R = \int\limits_{x \approx 1.6{\rm{ mm}}}^{x \approx 13{\rm{ mm}}} {r (x)dx} $, where $r(x)$ is the resistance per unit length given in (\ref{hyd_res_unit_length}). $QR$ vs. $\Delta p_e $ data for each channel and flow rate are shown in figure 4. The error bars are due to the propagated uncertainties in the measurements of $2h_0+\zeta(x)$.

\begin{figure}
\centering
\includegraphics[width=3.4in]{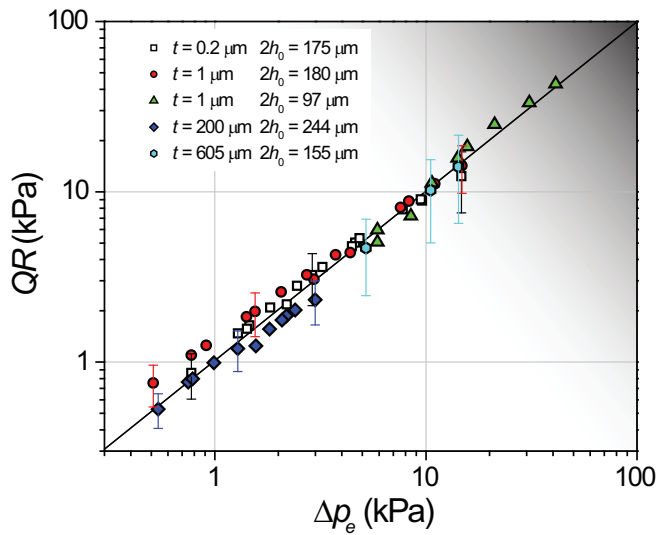}
\caption{ Calculated pressure drop $QR$ in the test section as a function of the extracted pressure drop $\Delta p_e$. The symbols match with those used in figure 2(a). Representative error bars are due to the uncertainties in $2h_0+\zeta(x)$. The solid line is the $QR=\Delta p_e$ line.}
\label{Fig4}
\end{figure}
\section{Conclusions and Outlook}
The agreement in figures 3 and 4 provides validation for our method and gives us confidence that we can measure $p(x)$ in  deformable channels accurately. For the proof-of-principle demonstration in this manuscript, we have applied our method to a flow, which can be approximated by the Poiseuille equation, i.e., (\ref{analytic_def_chnl}) and (\ref{hyd_res_unit_length}). However, the method should remain accurate independent of the nature of the flow (e.g., turbulent flows, flows with non-linear $p(x)$ or separated flows) because $p(x)$ simply comes from the wall response  ---  as evidenced by the \emph{non-linear} $p(x)$  resolvable in Fig  3(b).  It must also be re-emphasized that the nature of the wall response is not of consequence as long as the deflection is a continuous function --- any function --- of pressure, $\zeta=\zeta(p)$.  All these suggest that the method can be applied universally as an accurate probe of flows with micron, and even possibly sub-micron, length scales.

Our results may be related to prior studies on collapsible tubes \cite{Carpenter,Pedley-Carpenter}. For our system in two dimensions, in the case of small displacements,
 $\zeta/2h_{0}\ll1$, one  can write a ``tube law"
\begin{equation}
    p= p(\zeta, \zeta'')\approx a\zeta+b\zeta'',
\end{equation}
which relates the local the gauge pressure  $p$ (the so called transmural pressure) to the channel deflection $\zeta$ and its axial derivative  $\zeta''= d^{2}\zeta/dx^{2}$. In analogous expressions in the collapsible tube literature, the coefficient  $a$ is typically found by considering the changes in the cross-sectional area; however, finding  $b$, which determines the effect  of the axial tension on $p$,  is typically not simple and is possible only for certain tube geometries, e.g., for elliptic tubes \cite{Whittaker}.  Several points are noteworthy about our experiments. First, the axial tension term appears to be unimportant here, i.e., $p= p(\zeta)$. Second, the method remains accurate even when $\zeta \sim 2h_0$. Finally, a method similar to the one described here may be useful for determining $b$ experimentally for different geometries and large deformations.

We mention in passing that the friction drag in a channel with rigid walls separated by a gap of $2h_0$ is larger than that in a deformable channel with the same unperturbed gap. To see this, consider a one-dimensional flow with a flux $q$ per unit width. The stress at the rigid wall is $\tau  =  - \eta {\left. {{\partial_{y}u}} \right|_{{\rm{wall}}}} = {h_0} {{\partial_{x}p}}$. But $q={\frac{2 {h_{0}}^{3}}{3\eta }} {\partial_{x}p}$. Therefore, $\tau  = {{3{\eta ^2}q} \over {2{h_0}^2}}$. Given that ${q \over {{h_0}^2}} \ge {q \over {{{\left( {{h_0} + \zeta /2} \right)}^2}}}$, drag is reduced.  We also note that no evidence of transition to turbulence has been observed in our experiments even at the largest ${\rm{Re}}\approx 1200$.

This noninvasive method can possibly find applications in characterizing physiological flows. In blood flow in arteries \cite{Ku} and smaller vessels \cite{Popel},  flow-structure interactions are critical in determining functionality \cite{Grotberg,Pedley-Carpenter,HeilStokes}.  Using our method, for example, one could extract local pressure distribution in an arterial aneurysm, where the arterial wall  degrades and eventually ruptures due to the pressure and shear forces during blood flow \cite{Lasheras}.

The spatial resolution in a $p(x)$ measurement depends upon the resolution in $\zeta$, the noise in the hydrostatic pressure measurement, and the magnitude of the response of the wall. With our current imaging system, we can detect deflections with $\lesssim 20$ nm precision, and the r.m.s. noise in the hydrostatic pressure transducer is $\sim 10$ Pa. By collecting the constitutive curves in figure 3(a) at smaller pressure intervals, we estimate that we can measure $p(x)$ with $\sim10~\mu$m resolution in this particular channel. This method can easily be  scaled down to provide sub-micron resolution in a nano-fluidic channel by employing a higher numerical aperture objective. It may also be possible to extend the method to study time-dependent fluid-structure interactions \cite {Bertram, Huang} by collecting surface deformation maps faster \cite{Ashwin}. By optimizing the averaging time, one should be able to collect high-speed high-resolution pressure measurements  in miniaturized channels. Such  advances could open up many other interesting fluid dynamics problems, especially in biological systems.

\section{Acknowledgements}
We acknowledge generous support from the US NSF through Grant No. CMMI-0970071. We thank Mr. Le Li for help with the flow simulations.

\appendix
\section{}\label{appA}
Pressure drops in the rigid inlet and outlet regions of the channels are deduced from flow simulations in Comsol Multiphysics using the single-phase 3D steady laminar flow environment.  A constant volumetric flow rate is applied at the inlet port, and the pressure at the outlet port is kept at zero. All the channel walls are assigned the no-slip boundary condition. We use quadrilateral mesh elements and increase the mesh density until the results converge. The 2D simulations shown in figure 3(b) and (c) between the deformed top wall and the flat bottom wall are carried out using the same single-phase steady laminar flow environment. The upper compliant wall is replaced with a rigid wall, but the deformed  wall shape is preserved by importing the experimental profile into the simulation. At the inlet port, instead of volumetric flow rate, the calculated flow velocity corresponding to the experimental volumetric flow rate is used.

\end{document}